\parindent=0pt
\parskip=0.3cm
\magnification \magstephalf

\def \CM{{\hbox{\sevenrm CM}}}

\null
\vskip 1.5cm                 
\centerline{\bf COHERENT SCALAR FIELD OSCILLATIONS AND}
\centerline{\bf THE EPOCH OF DECELERATION}
\vskip 1.5cm

\centerline {by}
\vskip 0.3cm

\centerline {J.C. Jackson\footnote{*}{e-mail: john.jackson@unn.ac.uk}}
\centerline {School of Computing and Mathematics}
\centerline {University of Northumbria at Newcastle}
\centerline {Ellison Building}
\centerline {Newcastle upon Tyne NE1 8ST, UK}
\vskip 1.2cm

\centerline{\bf ABSTRACT}
\vskip 0.3cm

The discovery of supernova SN 1997ff at $z\sim 1.7$ has confirmed the 
expected switch from cosmological acceleration to deceleration, as 
predicted by the concordance $\Lambda$/CDM model.  However its position
in the SN Ia Hubble diagram suggests that the switch is too pronounced,
which here is taken to mean that a cosmological constant is not an
adequate description of the state of the vacuum.  An `oscillessence'
model is invoked, with a scalar field $\phi$ governed by a simple 
quadratic potential, which gives a better fit to the new data point.
The field is undergoing coherent oscillations, and a key feature of
the proposal is that we are towards the end of the second period of
acceleration; a $\Lambda/\phi$ mix replaces $\Lambda$/CDM, 
with $\Omega_\Lambda\sim 0.4$ and $\Omega_\phi\sim 0.6$.

\vskip 0.6cm

{\bf Key words:} cosmology -- observations -- theory -- dark matter. 
\vfil\eject

{\bf 1\hskip 0.3cm INTRODUCTION}

There is now a strong concensus that the basic cosmological parameters are
known, and that we are living in a spatially flat accelerating Universe,
with $\Omega_\CM\sim 0.3$ and $\Omega_\Lambda\sim 1-\Omega_\CM\sim 0.7$ (CM
$\equiv$ Cold Matter, that is Cold Dark Matter plus the baryonic component).
This concencus is based primarily upon observations of Type Ia supernovae 
(SNe Ia) (Schmidt et al. 1998; Riess et al. 1998; Perlmutter et al. 1999), 
coupled with observations of Cosmic Microwave Background (CMB) anisotropies  
(Efstathiou et al. 1999) and Large Scale Structure (LSS) information 
(Bridle et al. 1999; Lasenby, Bridle \& Hobson 2000; Efstathiou et al. 2001).
Separately these observations place different constraints on the parameter
values, but in combination degeneracies are removed.  However, the CMB
measurements have been refined through a number of ground-based and 
balloon-borne experiments (Halverson et al. 2001; Lee et al. 2001;
Netterfield et al. 2001), particularly significant being the detection
of multiple peaks in the CMB angular power spectrum, and the data sets 
are now beginning to tell the same story when considered separately
(Balbi et al. 2000; de Bernardis et al. 2001; Jaffe et al. 2001;
Pryke et al. 2001; Stompor et al. 2001).  Support for flatness comes
from measures of the angular size of the acoustic horizon at decoupling,
via detection of the first Doppler peak in the CMB angular spectrum
(de Bernardis et al. 2000; Hanany et al. 2000); for this reason I shall 
restrict my attention to spatially flat models.

This impressive convergence of independent measures of the various 
cosmological parameters is coming to be known as `concordance', and the 
corresponding cosmological model as the concordance model.  The concordance 
model predicts a transition from acceleration to deceleration at a redshift 
$z=(2\Omega_\Lambda/\Omega_\CM)^{1/3}-1\sim 0.67$, beyond which the matter 
density is dominant; the discovery of SN 1997ff (Gilliand \& Phillips 1998), 
together with fortuitous retrospective multicolour photometric observations 
which have established its type, a redshift $z\sim 1.7$ and 
a distance modulus, provide strong evidence that the epoch of deceleration 
has been observed (Riess et al. 2001).  I shall address here what appears 
to be a discordant note arising from these observations, which is a strong 
suggestion that the switch to deceleration is too violent.  Figure 1 is a 
Hubble diagram for SNe Ia, reproducing the results presented in 
Figure 11 of Riess et al. (2001), in which the points with error bars are 
redshift-binned data from Riess et al. (1998) and Perlmutter et al. (1999); 
here the magnitude $\Delta m$ is relative to an empty non-accelerating 
universe. The cross shows SN 1997ff at $z=1.7$, $\Delta m=-0.5$, 
and the corresponding 68\% and 95\% confidence regions are indicated.
The continuous curve is an unweighted least-squares fit to the $z<1$ points,
($\Omega_\CM=0.28$, $\Omega_\Lambda=0.72$), which clearly illustrates the 
point I am making here.  The dashed curve shows an unweighted least-squares
fit to all the points ($\Omega_\CM=0.44$, $\Omega_\Lambda=0.56$);
the extra matter produces more deceleration, but destroys the good fit 
to the $z<1$ points.  The possibility discussed here is that this 
behaviour is evidence that a simple cosmological constant is not an adequate 
description of the present state of the vacuum, and the purpose of this 
letter is to present a simple alternative which is compatible with the SN Ia
Hubble data, and might at the same time do least damage to the concordance 
picture.  
\vskip 0.3cm

{\bf 2\hskip 0.3cm $\Lambda/\phi$ MODELS}

Some time ago (before the case for acceleration was as strong as 
it now appears to be) (Jackson 1998a,b; Jackson \& Dodgson 1998;
see also Frieman et al. 1995) I considered models dominated by a homogeneous 
dynamical scalar field $\phi$ governed by an inflationary potential 
corresponding to an ultra-light inflaton:
$$
V(\phi)={\Lambda \over 8\pi G}+{1 \over 2}\omega_c^2\phi^2
\eqno(1)
$$

where $\omega_c$ is the associated Compton frequency, here taken to be 
somewhat larger than the inverse of the present Hubble time.  The scalar 
field has effective density $\rho_\phi=(\dot\phi^2+\omega_c^2\phi^2)/2$ and 
pressure $p_\phi=(\dot\phi^2-\omega_c^2\phi^2)/2$.  The most convenient 
formulation of the Friedmann equations governing the scale factor $R(t)$ 
is in terms of the instantaneous values of the deceleration parameter $q(t)$,
and the various density parameters $\Omega=8\pi G\rho/3H^2$; in the spatially 
flat case these are

$$
q=\Omega_\CM/2+(1+3w)\Omega_\phi/2-\Omega_\Lambda
\eqno(2)
$$
and
$$
\Omega_\CM+\Omega_\Lambda+\Omega_\phi=1
\eqno(3)
$$

where $w$ is the ratio $p_\phi/\rho_\phi$.

The scalar field is governed by the equation

$$
\ddot\phi+3H\dot\phi+\omega_c^2\phi=0,
\eqno(4)
$$

where $H=\dot R/R$ is Hubble's `constant'.  Initially 
(when $H>>\omega_c^{-1}$) the field $\phi$ undergoes a slow roll towards 
the minimum in $V(\phi)$ at $\phi=0$, during which phase the effective 
cosmological constant is $\Lambda+4\pi G\omega_c^2\phi^2$.  This is the 
standard inflationary picture, but in this context the pressure-free matter 
is assumed to be dynamically dominant during the slow-roll phase.  
Thereafter, when $H<\omega_c^{-1}$, $\phi$ undergoes coherent 
oscillations, and the speculation here (and in Jackson 1998a,b; Jackson
\& Dogson 1998) is that the Universe has entered a phase in which the scalar 
field is dynamically dominant and is executing such oscillations.  During 
this phase periods of violent deceleration (when $\rho_\phi+3p_\phi>0$)
alternate with periods of not-so-violent acceleration (when
$\rho_\phi+3p_\phi<0$), which behaviour can account for the SN 1997ff
observations.  Figure 2 is a typical example, showing the evolution of the
deceleration parameter $q(t)$ and the age/Hubble time ratio $t/t_H$.  In what
follows I shall assume generically that we are currently at point P, at the
end of the second period of acceleration, to maximise the effect of the
latter.  In the example shown the parameter values at P are
$\Omega_\CM=0.012$, $\Omega_\Lambda=0.433$, $\Omega_\phi=0.555$; their sum
is 1 as dictated by equation (3), and $q=0$ at P fixes $w$ (or equivalently
$\phi$ and $\dot\phi$) and hence the model according to equation (2). Locally
these values give the best fit to the SN Ia data, including SN 1997ff; 
the corresponding curve is shown in Figure 3.  Observation points located 
during the first period of acceleration do no better in this respect than 
straightforward $\Lambda/\hbox{CM}$ models, as in Figure 1.  Even later 
periods of acceleration might be considered, but the corresponding Hubble 
curves generally have features for which there is no evidence.

A interesting possibility is that just the quadratic term
in the potential (1) might suffice, with no true cosmological constant,
but this is discounted by the dashed curve in Figure 3, with $\Omega_\CM=0$,
$\Omega_\Lambda=0$, $\Omega_\phi=1$; deceleration sets in too soon in this
case, and addition of some Cold Matter would clearly make the situation
worse.  The picture which emerges here is that of a mixed $\Lambda/\phi$
model, with a low Cold Matter content, of the order of the baryon content
of the Universe (see Coc et al. (2001) for a recent review).  If SN Ia 
data were the only consideration then this model would be perfectly viable, 
but almost certainly not compatible with CMB and LSS.  However, there are 
acceptable models (i.e. when SN 1977ff is included) with more Cold 
Matter (but not more than ${\Omega_\CM}\sim 0.1$), which nevertheless 
probably are compatible.  This universe is heading for heat death, with
a final oscillatory flourish before oblivion; this is illustrated in
Figure 4.

I have allowed blatant fine-tuning here, with no attempt to account for
the balance between the two components of dark energy.  Following the 
seminal work of Ratra \& Peebles (1988) and Wetterich (1988), who considered
decaying potentials of the form $V(\phi)\propto \phi^{-\alpha}$ and
$V(\phi)\propto \exp(-\alpha\phi)$, the manufacture of scalar-field models 
has evolved into a large industrial concern, particularly post 1998 and
particularly with view to accounting for the balance between dark energy and 
dark matter in a natural way (see Bean and Melchiorri (2001) for a recent 
review).  Some of the corresponding potentials also engender oscillatory 
behaviour (for example Frieman et al. 1995; Skordis \& Albrecht 2000), 
and might serve as a alternatives to equation (1) in the present context.
Alternatively, if the simplicity of potential (1) is to be preferred, then
an anthropic argument might be found to account for the putative balance.
The generic term `oscillessence' might be reserved for dark energy which
allows multiple periods of acceleration/deceleration (cf. Caldwell, Dave 
\& Steinhardt 1998).
\vskip 0.6cm

{\bf FIGURE CAPTIONS}
\vskip 0.3cm

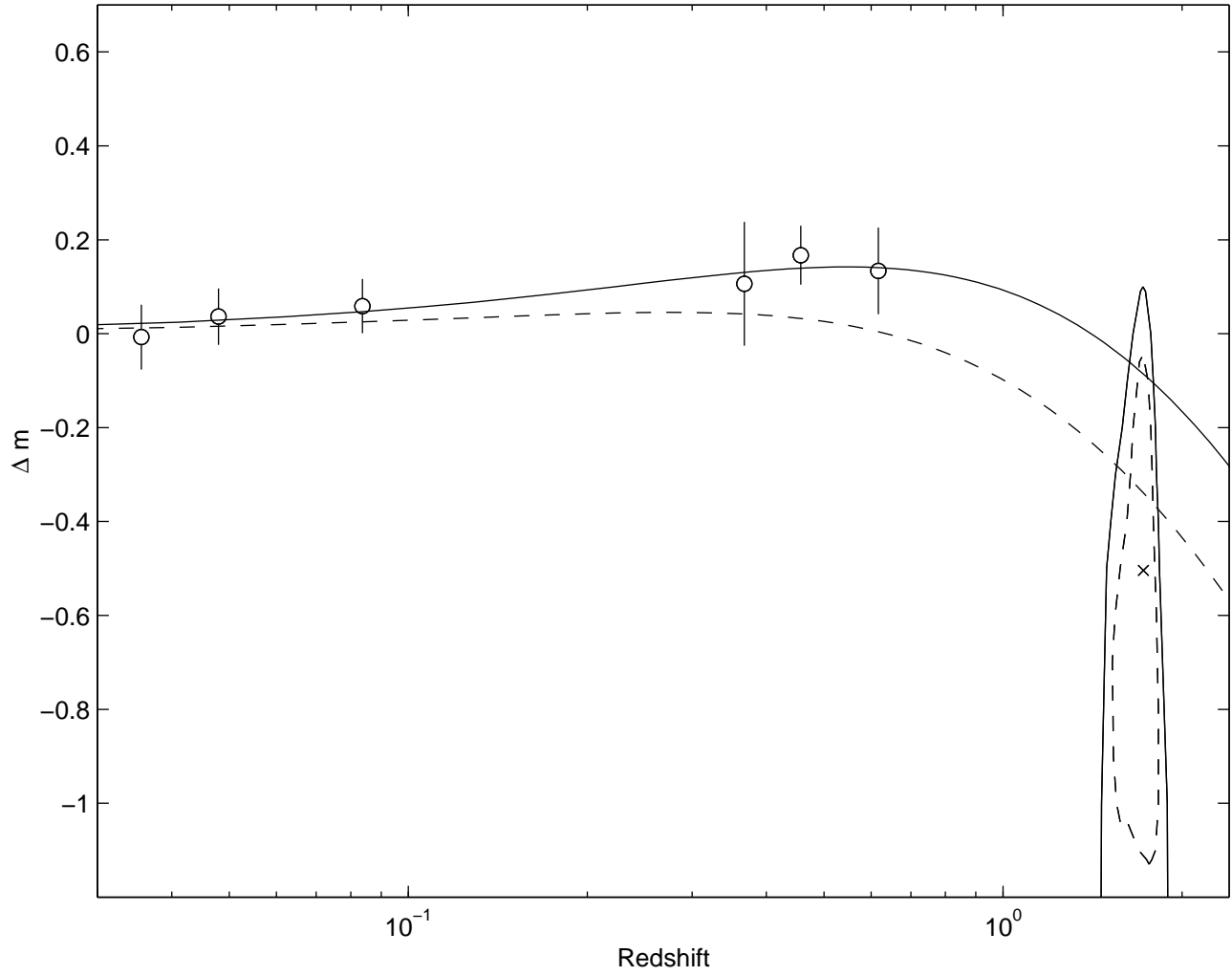
Figure 1.  Type 1a supernovae and $\Lambda$/CM models.  Hubble diagram 
relative to to an empty non-accelerating universe.  The cross indicates 
SN 1997ff, with 68\% and 95\% confidence regions.  The continuous curve
shows a concordance model ($\Omega_\CM=0.28$, $\Omega_\Lambda=0.72$); the
dashed curve is the best fit to all the data points ($\Omega_\CM=0.44$, 
$\Omega_\Lambda=0.56$).

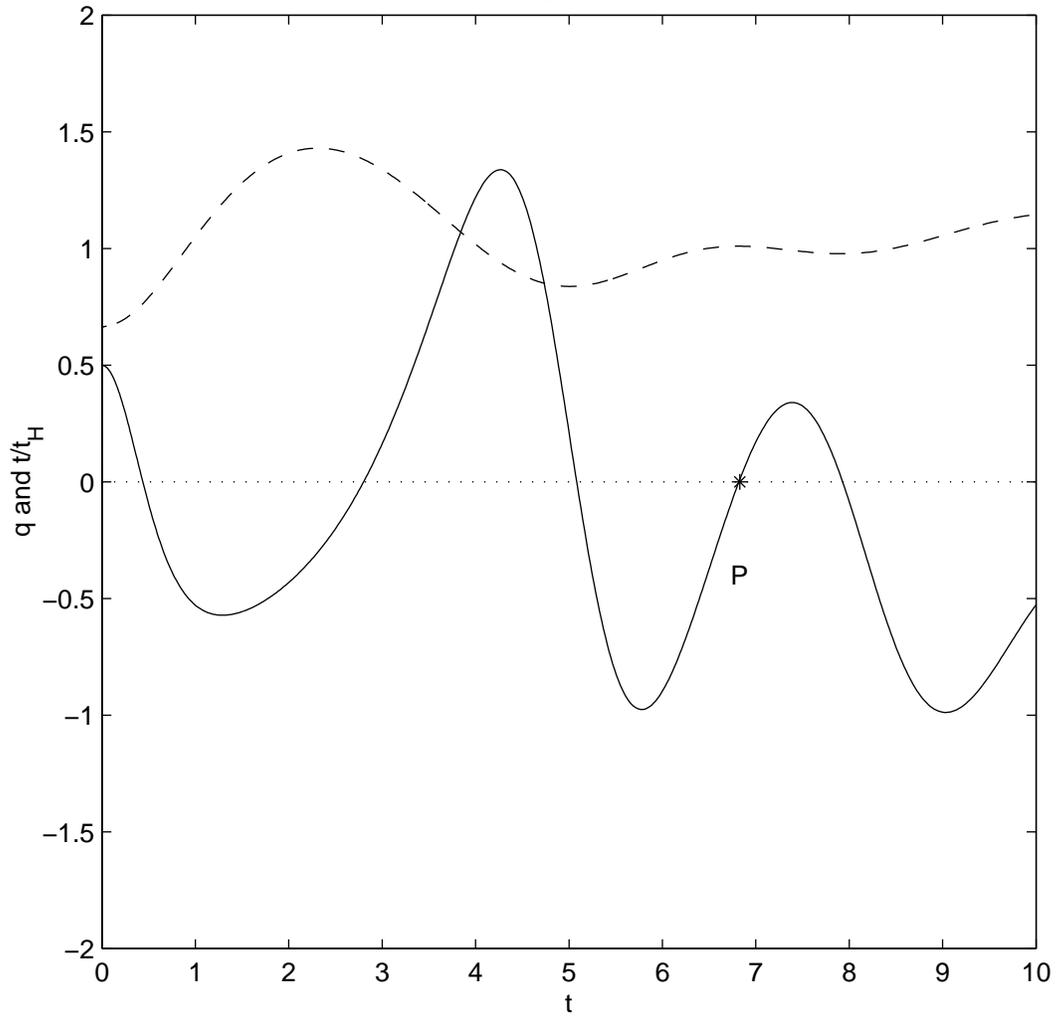
Figure 2.  An oscillating model, showing the evolution of the deceleration
parameter $q(t)$ (continuous) and the ratio $t/t_H$ (dashed).  The current
observation point is located at P, where $\Omega_\CM=0.012$, 
$\Omega_\Lambda=0.433$, $\Omega_\phi=0.555$.

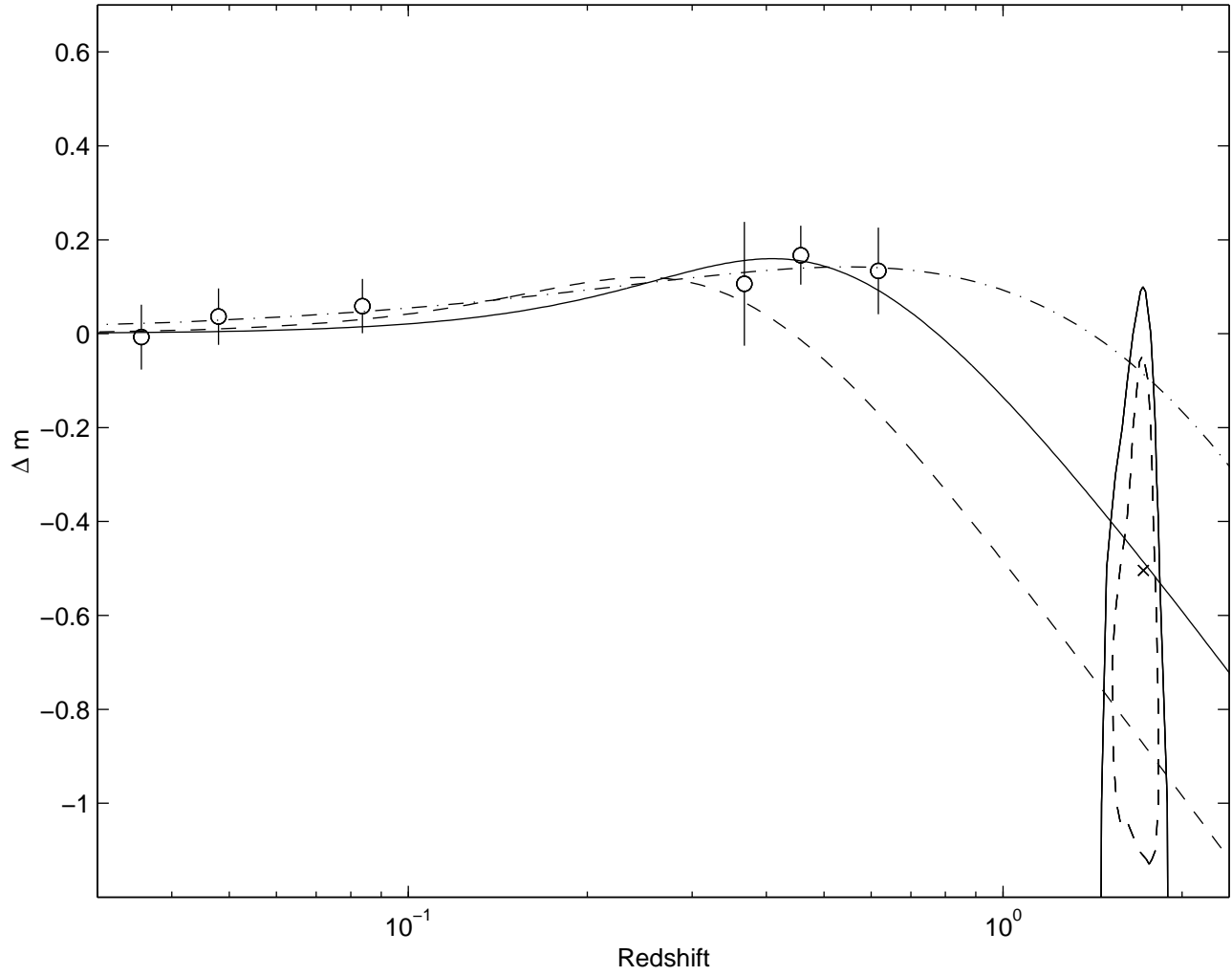
Figure 3.  Type 1a supernovae and $\Lambda/\phi$ models.  Hubble diagram 
relative to to an empty non-accelerating universe.  The cross indicates 
SN 1997ff, with 68\% and 95\% confidence regions.  The continuous curve
shows the best fitting $\Lambda/\phi$ model, with parameters as in Figure 2.  
The dashed curve has parameters $\Omega_\CM=0$, $\Omega_\Lambda=0$, 
$\Omega_\phi=1$.  The dash-dot curve is the concordance model as in 
Figure 1.

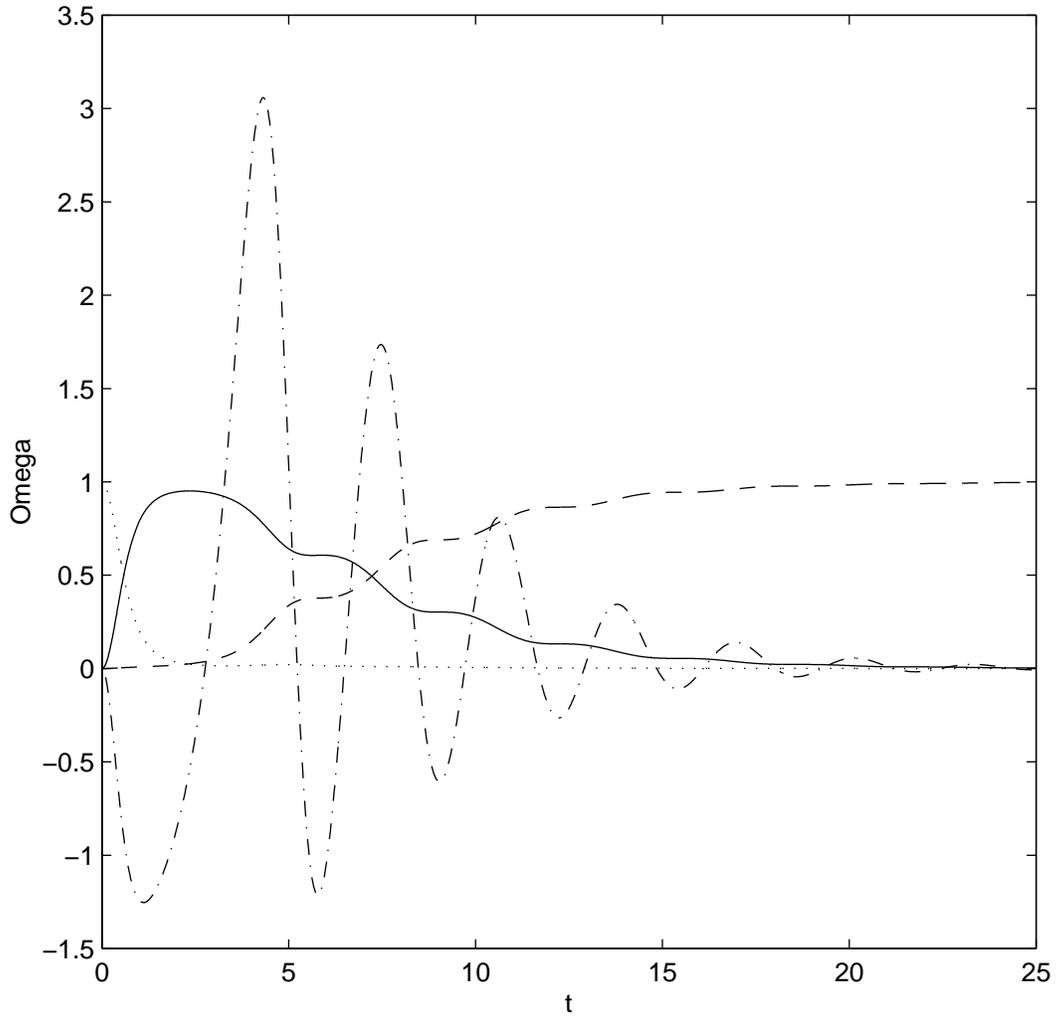
Figure 4.  Oscillating model showing the evolution of various density
parameters $\Omega=8\pi G\rho/3H^2$; $\Omega_\CM$ (dotted), $\Omega_\Lambda$
(dashed), $\Omega_\phi$ (continuous).  The dash-dot curve is a composite
parameter based upon the active gravitational density $\rho_\phi+3p_\phi$.
\vfil\eject

{\bf REFERENCES}

Skordis C., Albrecht A., 2000, astro-ph astro-ph/0012195

Balbi et al., 2001, ApJ, 2000, 545, L1; astro-ph/0005124

Bean R., Melchiorri A., 2001, astro-ph/0110472

Bridle S.L., Eke, V.R., Lahav O., Lasenby A.N., Hobson M.P., Cole S.,
Frenk C.S., Henry J.P., 1999, MNRAS, 310, 565

Caldwell R.R., Dave R., Steinhardt P.J., 1998, Phys. Rev. Lett., 80, 1582

Coc A., Vangioni-Flam E., Cass\'e M., Rabiet M., 2001, astro-ph/0111077

de Bernardis P. et al., 2000, Nature, 404, 955

de Bernardis P. et al., 2001, astro-ph/0105296

Efstathiou G., Bridle S.L., Lasenby A.N., Hobson M.P., Ellis R.S., 1999,
MNRAS, 303, 47

Efstathiou et al., 2001, astro-ph/0109152

Frieman J.A., Hill C.T., Stebbins A., Waga I., 1995,
Phys. Rev. Lett., 75, 2077
                                     
Gilliland R.L., Phillips M.M., 1998, IAU Circ. 6810

Hanany S. et al., 2000, ApJ Letters, 545, 5

Halverson N.W. et al., 2001, astro-ph/0104489

Jackson J.C., 1998a, MNRAS, 296, 619

Jackson J.C., 1998b, Mod. Phys. Lett. A, 13, 1737

Jackson J.C., Dodgson M., 1998, MNRAS, 297, 923

Jaffe et al., 2001, Phys. Rev. Lett., 86, 3475; astro-ph/0007333

Lee A. T. et al., 2001, astro-ph/0104459

Lasenby A.N., Bridle S.L., Hobson, M.P., 2000, 
Astrophys. Lett. \& Communications, 37, 327

Netterfield C.B. et al., 2001, astro-ph/0104460

Ratra B., Peebles P.J.E., 1988, Phys. Rev., D37, 3406

Perlmutter S. et al., 1999, ApJ, 517, 565

Pryke, C., 2001, astro-ph/0104490

Riess A.G. et al., 1998, AJ, 116, 1009

Riess A.G. et al., 2001, astro-ph/0104455

Schmidt B.P. et al., 1998, ApJ, 507, 46

Stompor et al., 2001, astro-ph/0106451

Wetterich, C., 1988, Nucl. Phys. B, 302, 668

\bye